\documentclass[12pt,a4paper]{article}
\usepackage{latexsym,amsmath,graphicx,epsfig,amssymb,rotating}
\usepackage{amssymb}

\newcommand{\mt}{t\kern-0.035cm\char39\kern-0.03cm}
\newcommand{\ml}{l\kern-0.035cm\char39\kern-0.03cm}
\newcommand{\md}{d\kern-0.035cm\char39\kern-0.03cm}
\newcommand{\mL}{L\kern-0.035cm\char39\kern-0.03cm}
\newcommand\chapter{\pagestyle{myheadings}}
\newtheorem{theo}{Theorem}

\newtheorem{coro}{Corollary}
\newtheorem{defi}{Definition}

\newtheorem{Example}{Example}

\begin{document}

\title{Berge equilibria in n-person 2-strategy games}
\author{Pawe{\l} Sawicki\thanks{Institute of Mathematics, University of Gda\'nsk, Poland; e-mail: p.a.sawicki@wp.pl}, Jaros{\l }aw Pykacz\thanks{Institute of Mathematics, University of Gda\'nsk, Poland; e-mail: pykacz@mat.ug.edu.pl}, Pawe{\l} Bytner\thanks{Institute of Mathematics, University of Gda\'nsk, Poland; e-mail: pawel.m.bytner@gmail.com} }
\date{}
\maketitle
\thispagestyle{empty}

\begin{abstract}

An algorithm for finding all Berge equilibria in the sense of Zhukovskii in n-person 2-strategy games in pure and mixed strategies is given. 


\end{abstract}
\section{Introduction}

The idea of a solution concept of a game  that nowadays is called {\it Berge equilibrium} ({\em in the sense of Zhukovskii}) was launched by French mathematician Claude Berge \cite{Ber57} in his book {\it Th\'{e}orie g\'{e}n\'{e}rale des jeux \`{a} n personnes}. The idea of Berge equilibrium is in a sense opposite to the idea of Nash equilibrium. While Nash equilibrium is based on egoism: each player aims to maximize his own payoff, Berge equilibrium is based on altruism: each player's aim is to maximize payoffs of all the other players, so when every player does so, everyone is better off. 

	As it was shown by Colman {\em et al.}  \cite{CKMT11}, in 2-person games there is a perfect symmetry between Nash and Berge equilibria: Nash equilibria in a game $G$ become Berge equilibria in a game $G'$ obtained from the game $G$ by interchanging players' payoffs, and vice versa. Actually, Colman {\it et al.} considered only equilibria in pure strategies, but it is straightforward to see that this result refers also to equilibria in mixed strategies. Therefore, by  the historic \cite{Nas50} theorem which assures that in any finite game there exists at least one Nash equilibrium -- in pure or mixed strategies, there exists also at least one Berge equilibrium in every 2-person finite game. 

The situation becomes different already in the case of the simplest 3-person games in which each player has only two pure strategies. In the MSc Thesis  written by Bytner \cite{Byt16} (see also Pykacz {\em et al.} \cite{PFB19}) an example of a 3-person 2-strategy game with no Berge equilibria at all, neither in pure nor in mixed strategies, was given.

The literature providing algorithms for finding Berge equilibria is very scarce.  Algorithms presented in Colman {\em et al.} \cite{CKMT11}, which allow to find Berge equilibria in n-person game by finding the common Nash equilibria in a family of associated 2-person games, allow to find Berge equilibria in pure strategies only. Also a simple algorithm presented in a paper by Corley and Kwain \cite{CK15}, although this paper is entiteled `An algorithm for computing all Berge equilibria', allows to find Berge equilibria in pure strategies only.

The aim of this paper is to provide an algorithm to find all Berge equilibria in completely mixed strategies, and also Berge equilibria of a `mixed type' (some players play pure and some play mixed strategies), in n-person, 2-strategy games. This, together with an algorithm by Corley and Kwain \cite{CK15} for finding Berge equilibria in finite games in pure strategies, yields a complete algorithm for finding all Berge equilibria in n-person 2-strategy games.  Although the obtained result is modest since it refers to the very narrow class of games, to the best of our knowledge it is the first algorithm for finding Berge equilibria in non-pure strategies in normal form games involving more than two players.

\section{Basic Notions and Definitions}
Noncooperative finite game in normal form is a triple:
\begin{equation}
G = \langle N, (S_i)_{i\in N}, (U_i)_{i\in N}\rangle ,
\end{equation}
where $N = \{1,\ldots ,n\}$ denotes the set of players, $S_i$ is a finite set of pure strategies of a player $i$, and $U_i$ is a function from $S = \prod_{i\in N}S_i $ into the set of real numbers that describes payoffs possible to obtain by the player $i$. Mixed strategy of the player $i$ is identified with a probability distribution defined on the set $S_i$ of his pure strategies. The set of all mixed strategies of the player $i$ is denoted $\widetilde{S}_i$. When at least one player chooses a completely mixed (i.e., non-pure) strategy, payoffs are understood as suitable expected values, and the set of real-valued functions they form, defined on $\widetilde{S} = \prod_{i\in N}\widetilde{S}_i $, will be denoted $(\widetilde{U}_i)_{i \in N}$.  We do not distinguish between a game and its mixed extension, and when we write {\it strategy} we mean a general mixed strategy, with pure strategies being special cases of mixed ones. Let ${\bf s} = (s_1,\ldots ,s_n) \in  \prod_{i\in N}\widetilde{S}_i $ be a {\it strategy profile}, then by ${\bf s}_{-i}$ we denote the {\it incomplete strategy profile} ${\bf s}_{-i} = (s_1,\ldots s_{i-1}, s_{i+1},\ldots ,s_n)\in \widetilde{S}_{-i} =  \prod_{j \neq i}\widetilde{S}_j $. By a small abuse of symbols we make an identification $(s_i , {\bf s}_{-i}) = {\bf s}$.

\begin{defi}
{\em A strategy profile ${\bf s}^* = (s^*_1, \ldots ,s^*_n)  \in \widetilde{S}$ is a {\em Berge equilibrium} (in the sense of Zhukovskii) of the game $G$ if:}
\begin{equation}
\forall i \in N, \quad  \forall {\bf s}_{-i} \in \widetilde{S}_{-i}, \quad \widetilde{U}_i(s^*_i,{ \bf s}_{-i}) \leq \widetilde{U}_i({\bf s}^*).
\end{equation}
\end{defi}
Let us compare this notion with the notion of Nash equilibrium:

\begin{defi}
{\em A strategy profile ${\bf s}^* = (s^*_1, \ldots ,s^*_n) \in \widetilde{S}$ is a {\em Nash equilibrium} of the game $G$ if:}
\begin{equation}
\forall i \in N, \quad  \forall s_{i} \in \widetilde{S}_{i}, \quad \widetilde{U}_i(s_i,{ \bf s}^*_{-i}) \leq \widetilde{U}_i({\bf s}^*).
\end{equation}
\end{defi}

By $2\times 2\times 2$ game we mean 3-person game in which each player has two pure strategies. Such games are defined by a pair of tri-matrices:
\begin{Example}
Let us consider the following $2\times 2\times 2$ game studied in \cite{Byt16}. Pure strategies of the first, the second, and the third player are denoted $A_1, A_2$; $B_1, B_2$; $C_1, C_2$, respectively.
\begin{equation}\label{game}
 C_{1}\colon~ \bordermatrix{& B_1 & B_2 \cr 
A_1 & (2,1,0) & (1,1,1) \cr 
A_2  & (2,0,1) & (1,0,2)} \quad  ~C_{2}\colon~ \bordermatrix{& B_1 & B_2 \cr 
A_1 & (1,2,0) & (0,2,1) \cr 
A_2  & (1,1,1) & (0,1,2)}.
\end{equation}
The left-hand matrix refers to the pure strategy $C_1$ of the third player, while the right-hand matrix refers to his pure strategy $C_2$. Let us note that this game is a very special one: None of the players has any possibility to influence his own payoff, no matter if he uses any of his pure or mixed strategies. On the contrary, his payoffs depend exclusively on the choices of the remaining players. This means that every, absolutely every, strategy profile, no matter whether consisting of pure or of completely mixed, or not completely mixed strategies, is a Nash equilibrium. In \cite{Byt16} it was shown that this game has no Berge equilibrium at all, neither in pure, nor in mixed strategies by using a specific method that will be described in a subsequent paper. In the present paper we shall show the same using our general algorithm.
\end{Example}

\section{An algorithm for finding all Berge equilibria in n-person 2-strategy games}

To make the presented algorithm complete we shall start with reminding Corley and Kwain \cite{CK15} algorithm for finding Berge equilibria in finite games in pure strategies.

\subsection{Berge equilibria in finite games in pure strategies}
Corley and Kwain \cite{CK15} based their algorithm on the notion of disappointment incurred by a player choosing a specific strategy $s_i \in S_i $.
\begin{defi}
{\em The disappointment incurred by a player $i$ choosing strategy $s_i$ while the other players choose strategies ${\bf s}_{-i}$ is the number}
\begin{equation}
d_i ({\bf s}) = d_i (s_i, {\bf s}_{-i}) = \max_{{\bf t}_{-i}\in S_{-i}}u_i (s_i, {\bf t}_{-i}) - u_i(s_i, {\bf s}_{-i}).
\end{equation}
\end{defi}

\noindent They noticed that this definition immediately yields the following theorem:
\begin{theo}
The pure strategy profile  ${\bf s}^*$ is a Berge equilibrium for the game $G$ if and only if the disappointment $d_i ({\bf s}^*) = 0$ for all $i \in N$.
\end{theo} 

It follows from this Theorem that at least in small games the most efficient way to find all Berge equilibria in pure strategies is to construct what Corley and Kwain in \cite{CK15} call {\em disappointment matrix}, i.e., a matrix obtained from the payoff matrix of a game by replacing all payoffs by respective disappointments. Then Berge equilibria in pure strategies are pure strategy profiles for which in disappontment matrix an entry is a null vector.

It is easy to check that disappointment matrix of a game studied in Example 1 looks as follows:
\begin{equation}\label{DMEx1}
 C_{1}\colon~ \bordermatrix{& B_1 & B_2 \cr 
A_1 & (0,1,2) & (1,1,1) \cr 
A_2  & (0,2,1) & (1,2,0)} \quad  ~C_{2}\colon~ \bordermatrix{& B_1 & B_2 \cr 
A_1 & (1,0,2) & (2,0,1) \cr 
A_2  & (1,1,1) & (2,1,0)}.
\end{equation}
Thus, we see that this game has no Berge equilibria in pure strategies.

\subsection{Berge equilibria in n-person 2-strategy games in completely mixed strategies}
We shall explain an algorithm for finding all Berge equilibria in completely mixed strategies in n-person 2-strategy games taking as an example 3-person 2-strategy game. Generalization of this method to games with bigger number of players is straightforward.

Let us denote, respectively, pure strategies of the first, the second, and the third player by  $A_1, A_2$; $B_1, B_2$; $C_1, C_2$, and let us denote the general mixed strategy profile by  ${\bf s}^* = (p,q,r)  \in \widetilde{S}$, where $p,q,r\in [0,1]$ are probabilities with which the first, the second, and the third payer, respectively, chooses his first pure strategy. Let us assume that Berge equilibrium in completely mixed startegies of such a game is a strategy profile ${\bf s}^* = (p,q,r)$, where $p,q,r\in (0,1)$. Let us consider the expected  payoff of the first player. Because of linearity of functions that yield expected values of payoffs we have:
\begin{eqnarray*}
\widetilde{U}_1 (p,q,r) & = & qr\widetilde{U}_1 (p,B_1,C_1) + q(1-r)\widetilde{U}_1 (p,B_1,C_2)\\ & + & (1-q)r\widetilde{U}_1 (p,B_2,C_1) +(1-q)(1-r)\widetilde{U}_1 (p,B_2,C_2),
\end{eqnarray*}
i.e., $\widetilde{U}_1 (p,q,r)$ is a convex combination of all possible payoffs to the first player when he plays his completely mixed strategy $p \in \widetilde{S}_1$ while the remaining players play all possible combinations of their pure strategies. 

Because we have assumed that the strategy profile ${\bf s}^* = (p,q,r)  \in \widetilde{S}$ is a Berge equilibrium, the following inequalities must be satisfied:
\begin{eqnarray}
\widetilde{U}_1 (p,q,r) & \geq & \widetilde{U}_1 (p,B_1,C_1),\\ \widetilde{U}_1 (p,q,r) & \geq & \widetilde{U}_1 (p,B_1,C_2),\\ \widetilde{U}_1 (p,q,r) & \geq & \widetilde{U}_1 (p,B_2,C_1),\\  \widetilde{U}_1 (p,q,r) & \geq & \widetilde{U}_1 (p,B_2,C_2).
\end{eqnarray}
\noindent But when a number is a non-trivial convex combination of other numbers and at the same time it is greater than or equal to each of them, all the considered numbers must be equal. Therefore, the probability $p$ must be a solution of the system of equations:
\begin{equation}
 \widetilde{U}_1 (p,B_1,C_1) =  \widetilde{U}_1 (p,B_1,C_2) =  \widetilde{U}_1 (p,B_2,C_1) =  \widetilde{U}_1 (p,B_2,C_2).
\end{equation}
\noindent For the same reasons the probability $q$ must be a solution of the system of equations:
\begin{equation}
 \widetilde{U}_2 (A_1,q,C_1) =  \widetilde{U}_2 (A_1,q,C_2) =  \widetilde{U}_2 (A_2,q,C_1) =  \widetilde{U}_2 (A_2,q,C_2),
\end{equation}
\noindent and the probability $r$  must be a solution of the system of equations:
\begin{equation}
 \widetilde{U}_3 (A_1,B_1,r) =  \widetilde{U}_3 (A_1,B_2,r) =  \widetilde{U}_3 (A_2,B_1,r) =  \widetilde{U}_3 (A_2,B_2,r).
\end{equation}

Of course in the general case of n-person 2-strategy game we obtain in this way n systems of equations, each system consisting of $2^{n-1}$ equations. Since in each of these systems of equations there is only one unknown variable, the bigger is the number of equations, the less probable is to find a solution. Therefore, we infere that the existence of Berge equilibria in completely mixed strategies in such games is rather an exception than a rule.
\begin{Example}
Let us consider the following $2\times 2\times 2$ game: 
\begin{equation}\label{game}
 C_{1}\colon~ \bordermatrix{& B_1 & B_2 \cr 
A_1 & (7,5,2) & (4,2,0) \cr 
A_2  & (3,7,-4) & (6,1,6)} \quad  ~C_{2}\colon~ \bordermatrix{& B_1 & B_2 \cr 
A_1 & (1,1,-3) & (8,4,0) \cr 
A_2  & (9,3,6) & (2,3,-9)}.
\end{equation}
It is easy to check that the systems of equations (11) -- (13) for this game look as follows:
\begin{equation}
4p+3 = -8p+9 = -2p+6 = 6p+2,
\end{equation}
\begin{equation}
3q+2 = -3q+4 = 6q+1 = 3,
\end{equation}
\begin{equation}
5r-3 = 0 = -10r+6 = 15r-9,
\end{equation}
\noindent and have the following unique solution:
\begin{equation}
p = 1/2, q = 1/3, r = 3/5.
\end{equation}
Therefore, the unique Berge equilibrium of this game in completely mixed strategies is the strategy profile  ${\bf s}^* = (1/2,1/3,3/5)$.
\end{Example}

Let us now find systems of equations (11) -- (13) for the game studied in Example 1. It is easy to check that in this case these systems of equations are as follows:
\begin{equation}
0=1=1=2,
\end{equation}
\begin{equation}
1=0=2=1,
\end{equation}
\begin{equation}
2=1=1=0.
\end{equation}
Since all these systems of equations are contradictory, we see that the game studied in Example 1 has no Berge equilibria in completely mixed strategies.

Let us note that in n-person, 2-strategy games all the considered equations are linear equations that have either none, or exactly one, or continuum of solutions, so we obtain the following corollary:
\begin{coro}
Any n-person, 2-strategy game has either none, or exactly one, or continuum of Berge equlilibria in completely mixed strategies.
\end{coro}

\subsection{Berge equilibria of a `mixed type' in n-person, 2-strategy games}
To make our paper complete, let us study now Berge equilibria of a `mixed type', i.e., such that some players play pure, while the remaining players play completely mixed strategies. Finding such equilibria in n-person, 2-strategy games is the most laborious task, nevertheless it is feasible. Fortunately enough, if a studied game has no equilibrium of this type, an algorithm finishes quickly.

Any equilibrium of a `mixed type' defines partition of the set of all players $N$ into two subsets: $N = P \cup M$ where the set $P$ consists of players that play pure strategies, and the set $M$ consists of remaining players that play completely mixed strategies. Of course if we are going to find all Berge equilibria of a `mixed type' we must check all possible partitions. In 3-player game there are 6 partitions, in 4-player game there are 14 partitions, and in n-player game there are $2^n-2$ partitions. (There are $2^n$ possibilities of choosing subset $P$ of set $N$ and subset $M$ is complementary to $P$ in $N$. Therefore $2^N$ partitions should exist, but two of them constitute special cases considered above.) This shows how laborious this algorithm is.

Now we shall assume the existence of strategy profile ${\bf b^*} = ({\bf p^*},{\bf m^*})$, where ${\bf p^*} = (p_1^*, p_2^*, \ldots, p_k^*) \in \prod_{i\in P}{S}_i$ and ${\bf m^*} = (m_1^*, m_2^*, \ldots, m_{n-k}^*) \in \prod_{i\in M}\widetilde{S}_i\setminus{S}_i$, forming a `mixed type' Berge equilibrium. Visibly, this can be true only if payoffs $\widetilde{U}_i$ of players in $P$ subset are the highest possible for their given strategies (on account of Definition 1). Moreover, as shown in subsection 3.2, payoff yielded by a given mixed strategy profile is a convex combination of all possible payoffs. It implies -- by the same reasoning as in subsection 3.2 -- that ${\bf b^*}$ may be a Berge equilibrium only if, for every pure strategy profile ${\bf m'}$ in $M$ and every player $p_i$, payoffs $\widetilde{U}_i$ for every strategy profile $({\bf p^*}, {\bf m'})$ are equal and maximal. Checking whether it really happens is straightforward (disappointment matrix), although fairly strenuous.

Number of subprofiles ${\bf p^*}$ fulfilling this condition is finite (just as number of {\it all} subprofiles ${\bf p}$), thus each of them may be investigated separately. On that basis we can assume that we already have a subprofile ${\bf p^*}$ with guaranteed maximal payoffs $\widetilde{U}_i$ for members of $P$ subset of players, needing only to ensure that members of $M$ subset will also get payoffs $\widetilde{U}_i$ maximal for their strategies (this could be named {\it Berge half-equilibrium}). Additionally, $M$ players now take part in a somehow self-contained game, as their payoffs rely only on their decisions. Thanks to this observation, we are now able to find all mixed Berge subequilibria (indeed, those are not {\it half-equilibria}, as we shall see soon) between them -- exploiting the algorithm given in subsection 3.2. If no such subequilibria exist, it becomes impossible for ${\bf b^*}$ to be a Berge equilibrium.

It has to be noted that listed conditions are not yet sufficient to ensure ${\bf m^*}$ being Berge half-equilibrium, and consequently ${\bf b^*}$ being Berge equilibrium, because (and only because) it could happen that some change in ${\bf p^*}$ increases the payoff $\widetilde{U}_i$ of some player in $M$. This possibility is easy to be confirmed or contradicted in finite number of steps when there is single subequilibrium in $M$, but things become complicated with continuum on them (as we already showed, there may be zero, one or continuum of such equilibria).

In order to complete the algorithm by solving this case, we begin with the observation: $\widetilde{U}_i(m_i,{ \bf m}_{-i},{\bf p^*})$, for a constant value of ${\bf p^*}$, is a linear function of $s_i \in (0, 1)$. It is of {\it crucial importance} that constancy of ${\bf m}_{-i}$ is not necessary, because -- as we already assumed -- subprofile ${\bf m}$ forms a Berge subequilibrium if complementary to the considered constant subprofile ${\bf p^*}$. Let us denote any strategy profile containing $m_i$ as $(m_i, {\bf b'}_{-i})$. Let us also denote any strategy profile composed of $m_i$ and {\it pure} strategies of remaining players as $(m_i, {\bf p'}_{-i})$. $\widetilde{U}_i(m_i,{ \bf p'}_{-i})$ is also a linear function of $s_i \in (0, 1)$ (and there exist $2^{n-1}$ such functions). As we already proved, every possible payoff $\widetilde{U}_i(m_i, {\bf b'}_{-i}))$ is a convex combination of payoffs $\widetilde{U}_i(m_i, {\bf p'}_{-i}))$. Therefore, $\widetilde{U}_i(m_i,{ \bf m}_{-i},{\bf p^*})$ is the highest of payoffs $\widetilde{U}_i(m_i, {\bf b'}_{-i}))$ if and only if it is not lower than any of payoffs $\widetilde{U}_i(m_i, {\bf p'}_{-i}))$.

However, it is actually possible to compare the linear function $\widetilde{U}_i(m_i,{ \bf m}_{-i},{\bf p^*})$ to the linear functions $\widetilde{U}_i(m_i, {\bf p'}_{-i}))$ one by one. Every such comparison consists of solving the linear inequality with one unknown and returns one of three results: `analysed function $\widetilde{U}_i(m_i, {\bf p'}_{-i}))$ never exceeds $\widetilde{U}_i(m_i,{ \bf m}_{-i},{\bf p^*})$ in $(0, 1)$ (and, as such, has no influence on Berge equilibria in question)', `analysed function $\widetilde{U}_i(m_i, {\bf p'}_{-i}))$ exceeds $\widetilde{U}_i(m_i,{ \bf m}_{-i},{\bf p^*})$ from $m_i=x$ onwards, resp. downwards (and, as such, limits Berge equilibria in question to $m_i \in (0, x]$, resp. $m_i \in [x, 1)$)' or `analysed function $\widetilde{U}_i(m_i, {\bf p'}_{-i}))$ exceeds $\widetilde{U}_i(m_i,{ \bf m}_{-i},{\bf p^*})$ all over $(0, 1)$ (and, as such, makes the existence of Berge equilibria in partition $N = P \cup M$ impossible)'. Having made $2^{n-1}$ such comparisons, in order to obtain a set of $m_i$ values participating in Berge equilibria in question it is enough to take the highest of lower limits and the lowest of upper limits. Obviously, these steps have to be repeated for every player $M_i$ having a continuum of strategies participating in Berge subequilibria in $M$. This completes the algorithm of finding Berge equilibria in n-person 2-strategy games.

For every partition of the players' set, the algorithm may be divided into three relatively simple steps. Step 1 is formed by finding the appropriate subprofiles ${\bf p^*}$. Step 2 contains solving, for each of found subprofiles, the problem of Berge subequilibria in $M$. Finally, step 3 consists of solving the aforementioned sets of linear inequalities.

\begin{Example}

Case study: First sportsman, Second sportsman and Trainer (FST) game. Trainer works in a tennis club in a small town and she may choose one of two forms of salary. In offer 1, she will get the same amount of money regardless of results of her trainees. In offer 2, her wage will be strongly dependant on these results. As for Sportsmen, they can choose `recreation' or `hard work'. Hard work always provides better payoff (in the sense of satisfaction); however, payoff resulting from hard work is higher with determined Trainer (offer 2), while payoff resulting from recreation is lower with determined Trainer, as then the training sessions are far too strenuous. Finally, we have to notice that Trainer, having chosen offer 2, prefers one devoted trainee than two, because in the latter case an immense amount of work outweighs the higher salary. These data allow us to write the payoff matrix:

\begin{equation}\label{FST}
 T_{1}\colon~ \bordermatrix{& S_1 & S_2 \cr 
F_1 & (2,2,2) & (2,3,2) \cr 
F_2  & (3,2,2) & (3,3,2)} \quad  ~T_{2}\colon~ \bordermatrix{& S_1 & S_2 \cr 
F_1 & (1,1,1) & (1,4,3) \cr 
F_2  & (4,1,3) & (4,4,2)}.
\end{equation}

For both Sportsmen, strategy 1 means recreational playing and strategy 2 means planning a tennis career. For Trainer, strategy 1 means taking offer 1 and strategy 2 means taking offer 2. It is easy to notice that the trainees have no influence on each other. Having the payoff matrix, we can construct the disappointment matrix:

\begin{equation}\label{FSTdis}
 T_{1}\colon~ \bordermatrix{& S_1 & S_2 \cr 
F_1 & (0,0,0) & (0,1,0) \cr 
F_2  & (1,0,0) & (1,1,0)} \quad  ~T_{2}\colon~ \bordermatrix{& S_1 & S_2 \cr 
F_1 & (1,1,2) & (1,0,0) \cr 
F_2  & (0,1,0) & (0,0,1)}.
\end{equation}

Therefore, we see that here exists only one Berge equilibrium in pure strategies. In order to find any completely mixed Berge equilibria, we have to construct three systems of equations like in Example 2:

\begin{equation}
-p+3 = -3p+4 = -p+3 = -3p+4,
\end{equation}
\begin{equation}
-q+3 = -3q+4 = -q+3 = -3q+4,
\end{equation}
\begin{equation}
r+1 = -r+3 = -r+3 = 2.
\end{equation}
\noindent One can easily check that they have no solution in $(0,1)^3$. Consequently, no Berge equilibria in completely mixed strategies exist here.

Our last task is to find all `mixed type' Berge equilibria. It is necessary to consider six possible partitions of the set of players: F and S pure -- T mixed; F pure -- S and T mixed; S pure -- F and T mixed; F and T pure -- S mixed; S and T pure -- F mixed; T pure -- F and S mixed.

In the case FS--T no pure strategy of F and S gives them equal and maximal payoff regardless of choice of T (as Trainer's attitude always influences satisfaction of Sportsmen). Therefore, the very first step of algorithm is enough to show that this partition gives no Berge equilibria. The same line of reasoning excludes F--ST and S--FT partitions.

FT--S is a bit more complicated. Step 1: the pairs of strategies $(F_1, T_1)$ and $(F_2, T_1)$ give F and T payoffs not influenced by S strategy. In spite of that, only the pair $(F_1, T_1)$ ensures that these payoffs will be maximal for the chosen strategy (which can be easily seen in the disappointment matrix: pair $(F_2, T_1)$ leaves F disappointed, as he hoped for $(F_2, T_2)$). Therefore, the algorithm continues with analysing just this one pair. Step 2: now we have to find all Berge subequilibria between $M$ players. Of course, we know how to do it in the general case. Moreover, it is obvious that every mixed strategy of S forms a Berge equilibrium in the subgame consisting {\it only of that player}. Step 3: the last thing to be done is solving three linear inequalities to determine the range of $q$ (we remind that $q$ describes the mixed strategy of S player). The linear function of Second sportsman's payoff for the strategy profile $(F_1, q, T_1)$ is $-q+3$. We have to compare it to the functions for profiles $(F_1, q, T_2)$, $(F_2, q, T_1)$, $(F_2, q, T_2)$. Two of these inequalities turn out redundant and the third one is:

\begin{equation}
-q+3 \ge -3q+4,
\end{equation}
\noindent which gives $q \ge 0.5$. Hence, the set of Berge equilibria for FT--S is $(p=1, 0.5 \le q<1, r=1)$. Because of full symmetry between First and Second sportsman, we also know that the set of Berge equilibria for ST--F is $(0.5 \le p<1, q=1, r=1)$.

The last case is T--FS. Step 1: only strategy $T_1$ gives T equal and maximal payoff regardless of F and S strategies. Step 2: every pair of mixed strategies of F and S forms a Berge subequilibrium between them, as they have no influence on each other's payoff (this can be also easily shown by applying the algorithm given in subsection 3.2). Step 3: the linear functions of First and Second sportsman's payoffs for the strategy profile $(p, q, T_1)$ are $-p+3$ and $-q+3$. We have to compare them to the functions for the profile $(p, q, T_2)$:

\begin{equation}
-p+3 \ge -3p+4; -q+3 \ge -3q+4,
\end{equation}
\noindent which gives $p, q \ge 0.5$. Hence, the set of Berge equilibria for T--FS is $(0.5 \le p<1, 0.5 \le q<1, r=1)$.

Summarizing the example, the set of all Berge equilibria in the FST game is $(p \ge 0.5, q \ge 0.5, r=1)$.

The interesting thing is that the set of all Nash equilibria in the FST game is $(p=0, q=0)$ (with no limits on r). The payoffs given by Berge equilibria are $\widetilde{U}_1, \widetilde{U}_2 \in [2, 2.5], \widetilde{U}_3=2$, whereas the payoffs given by Nash equilibria are $(4, 4, 2)$. The psychological reasoning behind the idea of Berge equilibrium, however, is getting the most one could hope for with his strategy (avoiding disappointment) rather than seeking for the highest payoff.

\end{Example}

\begin{Example}

For the sake of completeness, let us also finish the analysis of the game from Example 1. We have already solved cases with only pure strategies or only mixed strategies. Therefore, the six remaining cases are partitions AB--C, AC--B, BC--A, A--BC, B--AC, C--BA. It is instantly visible, however, that the consideration of each of these cases ends on step 1, because in all the configurations every player's payoff is dependant on decisions of his both rivals. This proves that this game has no Berge equilibria at all.

\end{Example}

\medskip

\noindent
\textbf{Acknowledgments.} Work by J. Pykacz was supported by the National Science Centre, Poland, under the research project 2016/23/D/ST1/01557.

\end{document}